# Development and characterization of a millimeter-wave cold load prototype

**Ruya Cong[1], Junjie Zhou[3], Yu Xu[2], Xuefeng Lu[2], Jianrong Cai[4], Haoxuan Gao[5], Kang Zhang[1], Xingyuan Hou[6], Zhengwei Li*[2] and Shaoliang Wang*[1]**

[1] *Institutes of Physical Science and Information Technology, Anhui University, Hefei 230601, China;*
[2] *State Key Laboratory of Particle Astrophysics, Institute of High Energy Physics, Chinese Academy of Sciences, Beijing 100049, China;*
[3] *School of Physics and Materials, Nanchang University, Nanchang 330031, China;*
[4] *School of Biomedical Engineering, Shanghai Jiao Tong University, Shanghai 200232, China;*
[5] *School of Mechatronics and Control Engineering, Shenzhen University, Shenzhen 518060, China;*
[6] *Center of Free Electron Laser and High Magnetic Field, Anhui University, Hefei 230601, China.*

*Correspondence: lizw@ihep.ac.cn

**Abstract:** Superconducting transition-edge sensors (TESs) are crucial detectors for cosmic microwave background (CMB) observations and require stable and tunable millimeter-wave cold loads for optical-efficiency calibration. This work presents the design, fabrication, and preliminary characterization of a 4–20 K millimeter-wave cold load prototype intended for integration into the 1 K stage of a dilution refrigerator and subsequent 40/90 GHz CMB TES calibration experiments. Two absorber prototypes based on commercially available CR-110 and a Stycast 2850FT composite were fabricated and studied. Simulation results show that both absorber structures exhibit small predicted steady-state temperature gradients and low normal-incidence reflectance in the target frequency bands. Room-temperature $S_{11}$ measurements were used only to screen low-reflectance cold load prototype, and the measured results generally agree with the electromagnetic simulations. The measured $S_{11}$ of the Stycast 2850FT composite is comparable to that of the commercial absorber TK RAM. Additionally, to explore a more readily obtainable alternative absorber material, TIE280-25AB was preliminarily evaluated by measuring its electromagnetic parameters. Based on the measured parameters, the simulated $S_{11}$ of the TIE280-25AB pyramidal absorber structure is comparable to those of CR-110 and the Stycast 2850FT composite over 33–110 GHz. These results identify CR-110 and the Stycast 2850FT composite as promising absorbers for subsequent cryogenic evaluation. The absolute low-temperature emissivity, effective radiation temperature, and TES calibration performance remain to be established through future cryogenic radiometric and TES based optical-power measurements.

## 1. INTRODUCTION

The B-mode polarization signal of the cosmic microwave background (CMB) provides crucial evidence for the existence of primordial gravitational waves [1, 2]. Current ground-based CMB polarization experiments widely use superconducting transition-edge sensors (TESs) arrays as the core focal-plane detectors [3]. These TES pixels are typically coupled to horn antennas and orthomode transducers (OMTs) to enable high-sensitivity polarization measurements [4, 5], with the primary observing bands centered at 90/150 GHz and an auxiliary 40 GHz band used to monitor and constrain low-frequency foregrounds such as Galactic synchrotron emission [6, 7]. As a type of bolometric detector, the TES requires traceable optical calibration to ensure the reliability of its scientific data, which establishes an accurate mapping between incident radiative power and electrical readout signals [8, 9]. Motivated by this requirement, this work develops and preliminarily characterizes a variable-temperature millimeter-wave cold load prototype for the 40/90 GHz bands. It is intended for future laboratory calibration of TES detectors, with potential application to the Ali CMB Polarization Telescope (AliCPT) [10, 11].

The intended cold load is designed to be installed on the 1 K stage of a dilution refrigerator (DR). Its target operating range is 4–20 K, which is intended to provide a low-background optical load while preventing overload of TES bolometers with picowatt-level saturation powers. For instance, the saturation powers of the Simons Observatory 90/150 GHz arrays are approximately 2.8/8.0 pW, respectively [12]. After complete cryogenic radiometric validation, an opaque cold load under thermal equilibrium can serve as a reference radiating surface with high emissivity. Its spectral shape follows Planck's law, while its total radiated power follows the Stefan–Boltzmann law ($P \propto T^4$) [13]. Consequently, by tuning the cold load temperature, the optical power incident on the TES can be varied continuously within a controlled range, enabling the calibration of key parameters, including optical efficiency [14, 15].

Structurally, the cold load features an oxygen-free high-conductivity copper (OFHC) backplate rigidly fixed to its bottom. This backplate provides mechanical support, blocks optical transmission, and serves as a standardized interface for mounting thermometers and heaters. Because the copper backplate suppresses transmission, low total reflectance is a necessary condition for high absorptance. Therefore, room-temperature normal-incidence $S_{11}$ is used in this work as a preliminary screening metric for selecting promising absorber, rather than as a substitute for direct absorptance or radiation-power measurements. A similar 1 K-class variable-temperature cold load has been demonstrated on the CMB-S4 single-module test platform [8]. This indicates that a variable-temperature cryogenic cold load is an established technical route for superconducting millimeter-wave detector characterization. A dedicated cryogenic measurement platform based on a 40 GHz TES detector system is currently under construction for subsequent evaluation of the prototype. From the perspective of practical implementation, established commercial absorber materials such as CR-110 may involve procurement uncertainty, long lead times, and relatively high cost. These considerations motivated us to evaluate TIE280-25AB from China as an alternative material and to develop a reproducible fabrication process.

This work presents the development and preliminary characterization of a millimeter-wave cold load prototype from four aspects: material characterization, simulation-driven design, fabrication, and room-temperature reflectivity evaluation. First, we measure the electromagnetic parameters of CR-110, the Stycast 2850FT composite, and the TIE280-25AB composite. For CR-110 and the Stycast 2850FT composite, we further analyze their cryogenic thermal properties and derive relevant models. Second, using the measured parameters, we establish full-wave electromagnetic and steady-state thermal models of four-sided pyramidal absorber structures. These models help determine the key dimensions and thermal design parameters required for low reflectance and good thermal stability in the target bands. The TIE280-25AB composite mainly serves as a comparison to evaluate materials. Third, we optimize a reproducible silicone replica-molding process to solve key fabrication challenges, including bubble entrapment, incomplete tip formation, and demolding-induced damage. Notably, the Stycast 2850FT composite, derived from the Berkeley Black formulation, is typically used as a wall coating. We therefore conducted cryogenic thermal-cycling tests on its cast components and confirmed that no cracking occurred at low temperatures. Finally, a vector network analyzer (VNA)-based $S_{11}$ measurement system is established to evaluate the room-temperature normal-incidence reflectivity of the fabricated absorber prototypes and to determine whether they satisfy the preset low-reflectance screening criterion. The results identify CR-110 and the Stycast 2850FT composite as promising absorbers for subsequent cryogenic radiometric and TES-based optical-power measurements, while TIE280-25AB remains a preliminary material requiring further fabrication and experimental evaluation.

## 2. DESIGN, SIMULATION AND FABRICATION OF COLD LOAD PROTOTYPE

To calibrate a TES detector in the laboratory, a reference source capable of delivering a controllable and repeatable radiative load under cryogenic conditions is required. In the millimeter-wave regime, the Rayleigh–Jeans law gives the optical efficiency $\eta_{\mathrm{optical}}$ for a single polarization and a single mode as [15–17]

$$\eta_{\mathrm{optical}} = \frac{P_{\mathrm{TES-2}} - P_{\mathrm{TES-1}}}{(T_2 - T_1)\, k_{\mathrm{B}} \int F(\nu)\, \mathrm{d}\nu} \tag{1}$$

where $P_{\mathrm{TES-1}}$ and $P_{\mathrm{TES-2}}$ are the TES powers measured at temperatures $T_1$ and $T_2$, respectively; $k_{\mathrm{B}}$ is the Boltzmann constant; and $F(\nu)$ is the normalized spectral response of the OMT. It follows that, by varying the cold load temperature, the incident optical power can be tuned continuously, thereby providing multiple known optical-loading points from which key parameters such as optical efficiency, responsivity, and saturation power can be inferred. Meanwhile, temperature drift translates directly into uncertainty in the applied optical load and degrades the fitting accuracy of the calibration. Therefore, the intended cold load should maintain stable temperature control and a small temperature gradient across the absorbing surface.

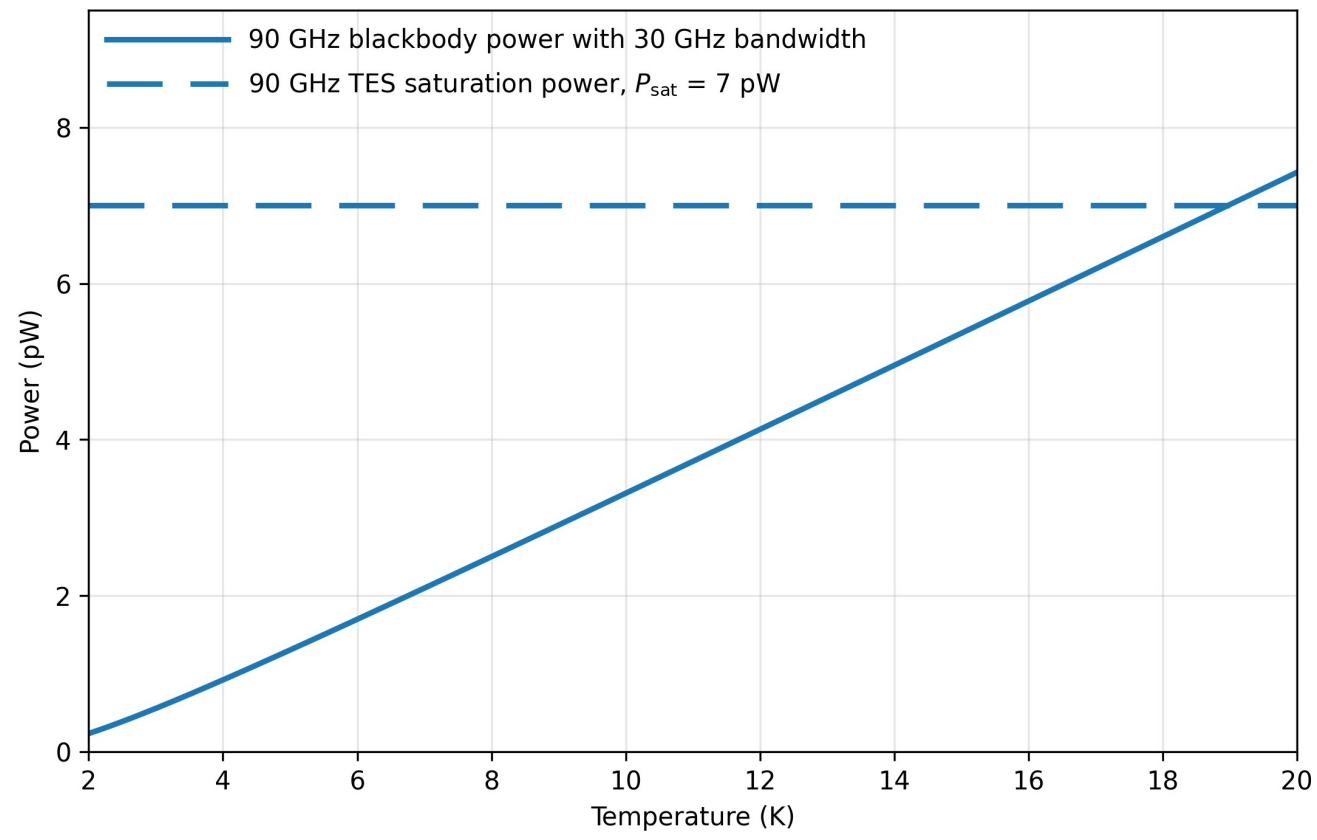

**Fig. 1: Temperature dependence of the ideal in-band optical power of a 90 GHz cold load. The calculation assumes a single spatial mode, a single polarization, and an effective bandwidth of 30 GHz. The dashed line shows the typical saturation power of a CMB TES detector.**

To evaluate the practical relevance of this temperature range, the ideal in-band optical power of the 90 GHz channel was calculated over 2–20 K using Planck's law. In this calculation, a single spatial mode, a single polarization, and an effective bandwidth of 30 GHz were assumed. As shown in Fig. 1, the calculated power spans a continuously tunable picowatt-level range and approaches the adopted 90 GHz TES saturation power of 7 pW [10] near 20 K. In practice, the usable temperature range should be chosen according to the actual optical coupling and TES saturation margin. This section addresses these requirements and the spatial constraints imposed by the DR, and presents absorber designs for a variable-temperature cold load. Specifically, the electromagnetic and cryogenic thermal parameters of absorbing materials are first characterized to provide inputs for subsequent modeling. Full-wave electromagnetic and thermal simulations are then conducted to determine the critical dimensions and thermal-control configuration. Finally, the room-temperature measurements are used to screen the absorber structures.

### 2.1 Characterization of absorber materials

To screen absorber materials for a TES cold load, the candidates should exhibit low reflectance in the target millimeter-wave bands and suitable thermal performance over 4–20 K. Accordingly, this section characterizes absorber materials from both electromagnetic and thermal perspectives, providing input parameters for the subsequent electromagnetic and heat-transfer simulations.

Considering material availability, machinability, and prior application experience, this study selects Eccosorb CR-110 [15], a Stycast 2850FT composite, and a TIE280-25AB composite as the absorbing materials. CR-110 is a two-part epoxy resin loaded with iron powder. It can be used directly without additional conductive fillers and retains strong absorption above 26 GHz. The Stycast 2850FT composite is derived from the Berkeley Black formulation system [18, 19], consisting of 68% Stycast 2850FT, 5% Catalyst 24LV curing agent, 5% carbon black, and 22% glass microbeads by mass. In this composite, carbon black is introduced to enhance the dielectric loss of the composite, while glass microbeads can improve the surface morphology of the composite and enhance electromagnetic-wave scattering to a certain extent, as summarized in Table 1.

To explore alternative resin materials for millimeter-wave cold loads, this study adopts a formulation design approach similar to that of Berkeley Black and prepares a TIE280-25AB composite absorber by completely replacing Stycast 2850FT with TIE280-25AB. TIE280-25AB not only exhibits a high dielectric constant, but also shows higher room-temperature thermal conductivity than CR-110 and Stycast 2850FT. Based on the above material systems, the subsequent sections first present the methods for measuring the electromagnetic parameters of all three materials, and then present the cryogenic thermal-parameter analysis and engineering-oriented derivation for CR-110 and Stycast 2850FT composites. In this work, TIE280-25AB is introduced as a preliminary alternative absorber material. Its electromagnetic parameters are measured and used for full-wave simulation, while prototype fabrication and low-temperature evaluation will be carried out in future work.

Table 1: Basic information of the absorbing materials investigated in this work

| Material | Main material source / country | Composition and density |
|---|---|---|
| CR-110 | Eccosorb CR-110, Laird Performance Materials / USA | Two-part castable epoxy-based microwave absorber loaded with iron powder, ($1.6\ \mathrm{g\,cm^{-3}}$). |
| Stycast 2850FT composite | Stycast 2850FT, Henkel / Germany | 68% Stycast 2850FT $\left(2.4\ \mathrm{g\,cm^{-3}}\right)$,<br>5% Catalyst 24LV curing agent $\left(1.02\ \mathrm{g\,cm^{-3}}\right)$,<br>5% carbon black $\left(2.3\ \mathrm{g\,cm^{-3}}\right)$,<br>22% glass microbeads $\left(0.125\ \mathrm{g\,cm^{-3}}\right)$. |
| TIE280-25AB composite | TIE280-25AB, Ziitek / China | 73% TIE280-25AB $\left(2.1\ \mathrm{g\,cm^{-3}}\right)$,<br>5% carbon black $\left(2.3\ \mathrm{g\,cm^{-3}}\right)$,<br>22% glass microbeads $\left(0.125\ \mathrm{g\,cm^{-3}}\right)$ |

#### 2.1.1 Electromagnetic Parameter Measurements

In the electromagnetic design of cold load absorbers, the complex permittivity and permeability directly affect interface reflection and energy dissipation and therefore constitute key inputs for full-wave simulations. In this work, absorber materials are characterized using a rectangular-waveguide method in the single-mode frequency range of 26.5–40 GHz. The scattering parameters of the samples are measured with a VNA and subsequently inverted to retrieve their electromagnetic parameters [20, 21].

The basic information of the three absorbing materials investigated in this work is summarized in Table 1. During measurements, each sample must fit tightly into the rectangular-waveguide cross section. Even small gaps between the sample and the metallic walls can compromise measurement repeatability and inversion stability. To eliminate such gaps, all samples are machined using a customized computer numerical control (CNC) process that precisely matches the waveguide dimensions and assembly tolerances. This achieves a high degree of conformity to the waveguide cavity while preserving the structural integrity of the materials. After fabrication, the key external dimensions and the assembly fit are inspected to ensure consistent mounting and comparable measurement results, as shown in Fig. 2A– 2C.

Fig. 2D– 2G presents the measured relative dielectric permittivity $\varepsilon_r = \varepsilon' - j\varepsilon''$ and relative magnetic permeability $\mu_r = \mu' - j\mu''$ of the three materials over 26.5–40 GHz. The extracted $\varepsilon_r$ and $\mu_r$ serve as key inputs for subsequent full-wave electromagnetic simulations to evaluate the reflectivity of the absorber structures in the 40/90 GHz bands. For CR-110, the measured $\varepsilon_r$ and $\mu_r$ are in good agreement with the typical ranges reported in the literature [22, 23], indicating that the sample fabrication, mounting, calibration, and parameter-extraction procedures used in this work provide good accuracy and reproducibility.

#### 2.1.2 Derivation of Cryogenic Thermal Properties

Evaluating the thermal feasibility of the cold load absorbers over 4–20 K requires estimates of their low-temperature thermal conductivity and specific heat capacity. For CR-110, the available cryogenic thermal-conductivity data are limited to a narrow range near 4 K, where the value is approximately 0.08 $\mathrm{W\,m^{-1}\,K^{-1}}$. No direct data exist at 20 K; we therefore adopted 0.26 $\mathrm{W\,m^{-1}\,K^{-1}}$ as an engineering estimate at that temperature. The density of CR-110 is about 1.6 $\mathrm{g\,cm^{-3}}$, and its specific heat capacity follows $C_{p,\mathrm{CR\text{-}110}} = 0.6T^{2.05}$ [24, 25].

By contrast, systematic cryogenic thermal data for the Stycast 2850FT composite remain scarce. This work therefore estimates its thermal properties for an engineering feasibility analysis. In the absence of direct measurements, we employ the Maxwell–Eucken (ME) model to estimate the effective thermal conductivity $k_{\mathrm{mix}}$ of the composite [28]. The ME model originates from Maxwell's treatment of an effective conductivity obtained by solving the Laplace equation; Eucken replaced the electrical conductivity in that framework with thermal conductivity, yielding a classical theoretical model applicable to composite materials. The model assumes that filler particles are spherical and uniformly dispersed in a continuous matrix, while neglecting interactions among the filler particles. The ME model is expressed as

$$k_{\mathrm{mix}} = k_m \frac{k_i + 2k_m + 2\phi\,(k_i - k_m)}{k_i + 2k_m - \phi\,(k_i - k_m)} \tag{2}$$

where $k_m$ is the thermal conductivity of the matrix, $k_i$ is the thermal conductivity of the filler, and $\phi$ is the filler volume fraction. The thermal conductivity of Stycast 2850FT in the range of 4–30 K, $k_{2850}$, has been reported by Koettig et al. [26]. Because the thermal conductivity and specific heat capacity of carbon black and glass microbeads are relatively small, they are assumed to be zero in the derivation; this assumption provides a conservative lower bound for the simulation results.

Given that the composite is a multi-filler system and that the morphology of the carbon-black particles does not fully conform to the ideal spherical-particle assumption, the Maxwell–Eucken (ME) model is used in this work primarily to obtain an engineering approximation of the effective thermal conductivity. The densities of the constituent materials are listed in Table 1.

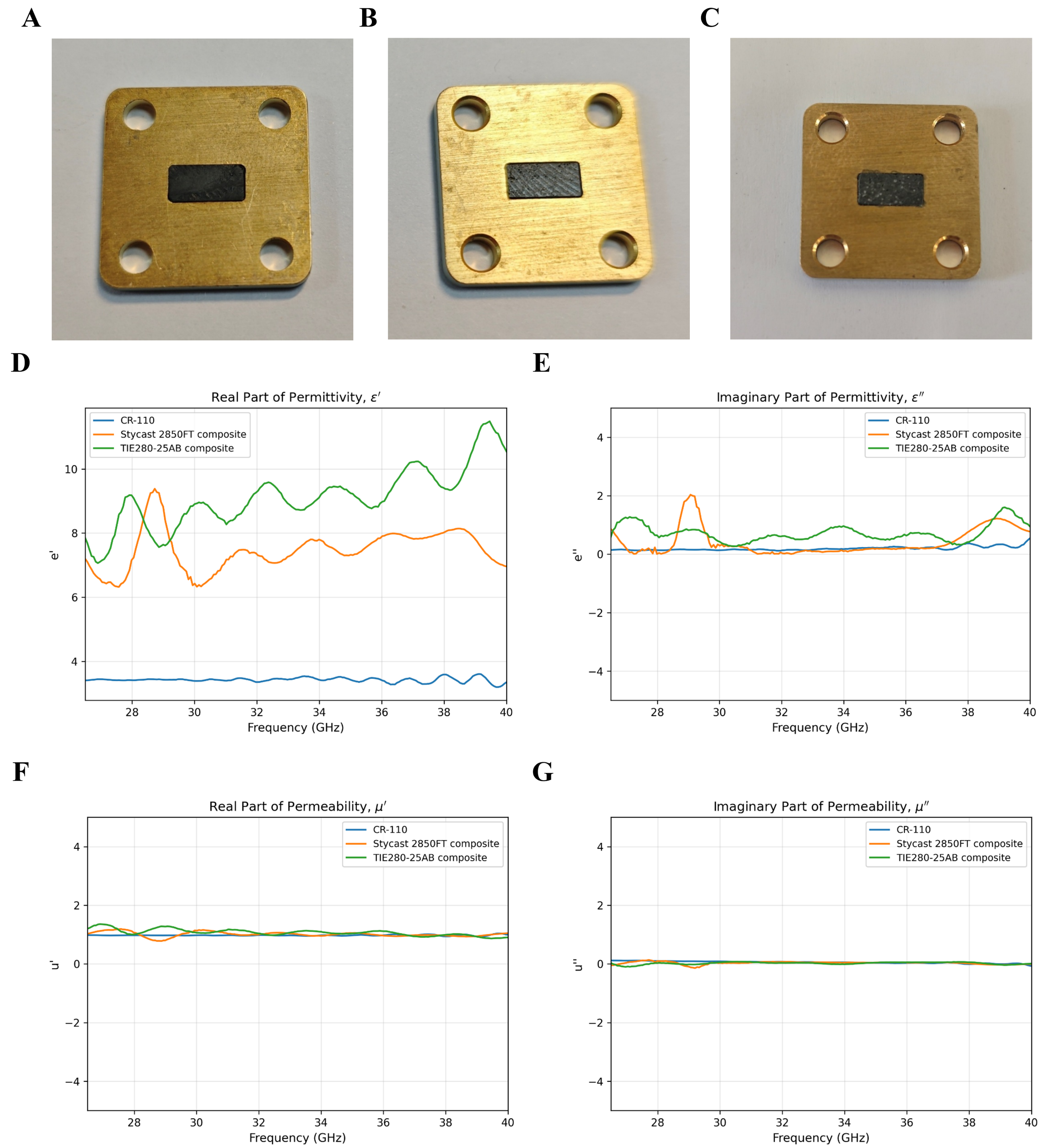

**Fig. 2: Electromagnetic characterization of the absorber materials. (A) CR-110. (B) Stycast 2850FT composite. (C) TIE280-25AB composite. (D) Real part of relative permittivity, $\varepsilon'$. (E) Imaginary part of relative permittivity, $\varepsilon''$. (F) Real part of relative permeability, $\mu'$. (G) Imaginary part of relative permeability, $\mu''$.**

It follows that $k_{\text{mix}} = 0.11k_{2850}$, and the overall density $\rho_{\text{mix}}$ is 0.473 $\text{g cm}^{-3}$.

The relationship between the specific heat capacity of Stycast 2850FT, $C_{p,2850}$, and temperature $T$ over 2–20 K

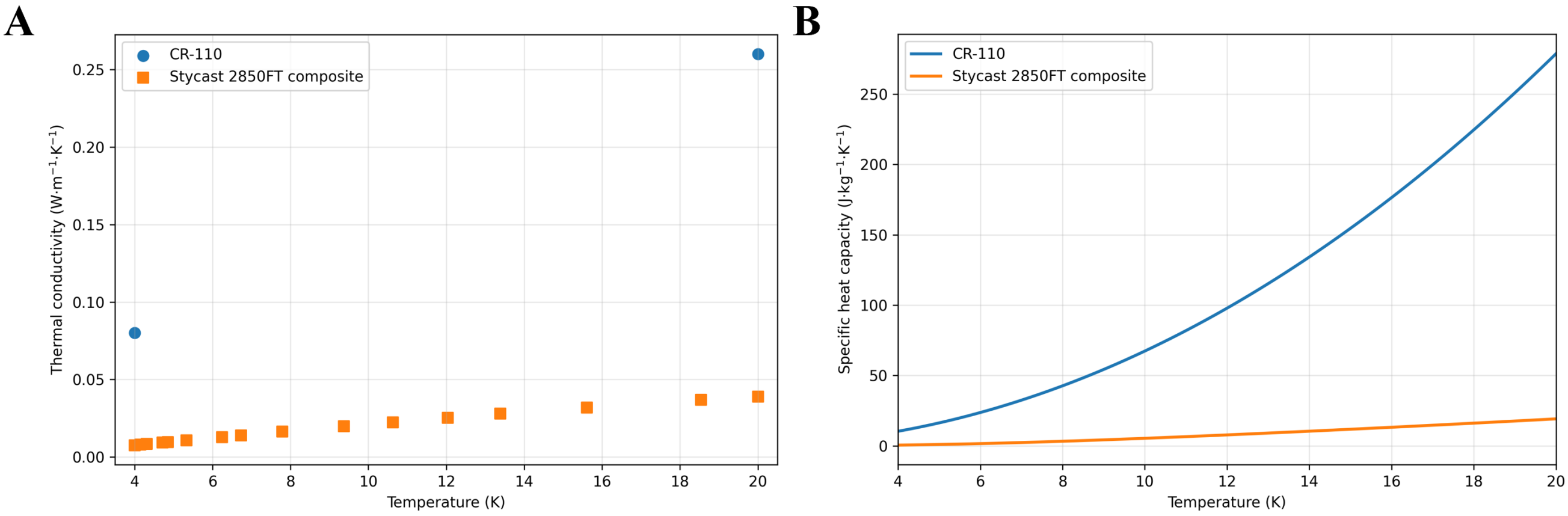

**Fig. 3: Estimated cryogenic parameters of CR-110[24, 25] and the Stycast 2850FT composite[26, 27] used in the heat transfer simulations. (A) Thermal conductivity. (B) Specific heat capacity. The plotted values are engineering estimates adopted for simulation input rather than directly measured properties of the fabricated absorber samples.**

is given by [27]

$$\log_{10} C_{p,2850} = \sum_{i=0}^{4} a_i \left(\log_{10} T\right)^i \tag{3}$$

where $a_0 = -2.11494050$, $a_1 = 2.51709056$, $a_2 = 2.59374465$, $a_3 = -2.99675188$, and $a_4 = 0.86125486$. Accordingly, $C_{p,\mathrm{mix}} = 0.73 C_{p,2850}$.

The engineering estimates of the cryogenic thermophysical parameters of CR-110 and the Stycast 2850FT composite used as inputs for the heat-transfer simulations are shown in Fig. 3.

TIE280-25AB has a higher nominal room-temperature thermal conductivity $\left(2.5\ \mathrm{W\,m^{-1}\,K^{-1}}\right)$ than CR-110 and Stycast 2850FT. However, its cryogenic thermal properties have not yet been characterized. In this work, the TIE280-25AB composite is evaluated only based on its electromagnetic performance and is not included in the present heat-transfer analysis.

## 2.2 Simulation of cold-load absorbers

### *2.2.1 Electromagnetic Simulation*

The DR offers only limited internal space, yet the intended cold load must provide an absorbing aperture large enough to cover the horn-antenna beam. To meet both constraints, we adopt a pyramidal absorber structure. The pyramid geometry creates a gradual impedance transition from free space to a lossy medium. Multiple reflections increase the wave–material interaction path, allowing low reflectance within a limited axial height. It can also be tiled laterally to accommodate the calibration of large-aperture TES arrays. For integration into the calibration system, the TES and horn-antenna assembly is mounted on the mixing-chamber (MXC) cold plate of the DR, whereas the absorber structure is fixed to the 1 K radiation shield by means of a low-thermal-conductivity G10 support to reduce the thermal load. To suppress out-of-band radiation and stray light, a metal-mesh optical filter is installed between the absorber structure and the horn antenna. In addition, a thermometer and heaters are incorporated to enable temperature control of the absorber structure and monitoring of key thermal nodes. The assembly configuration is shown in Fig. 4A. We conducted frequency-domain full-wave electromagnetic simulations of the pyramidal array to evaluate its reflective performance in the target millimeter-wave bands and to guide the choice of structural dimensions. The absorbing unit is a square-based, four-sided pyramid. This geometry offers good symmetry, which helps suppress cross-polarization [29], and promotes a more uniform temperature distribution across the absorbing surface during subsequent thermal design [30]. The unit is defined by two parameters: the base side length $L$ and the height $H$. The absorber structure must fit within the limited space of the 1 K shield in the DR. We therefore adopt $H/L = 3$ as the baseline design [29] to balance low reflectance against volume constraints. This aspect ratio represents a favorable compromise between reflectivity reduction and space occupancy. A larger $H/L$ could further reduce reflectance but would significantly increase the structural height and complicate spatial integration.

The core operating bands of the absorber structure are 40 GHz and 90 GHz, corresponding to free-space wavelengths $\lambda_0$ of approximately 7.5 mm and 3.3 mm, respectively. Under this condition, when the geometric-optics, or ray-tracing, method is adopted for simulation, the characteristic dimension of the absorber unit is on the same order of magnitude as the operating wavelength, and the electromagnetic response of the array is significantly affected by wave effects such as diffraction, interference, surface waves, and inter-unit coupling [31, 32]. Therefore, this simulation method is not applicable to the millimeter-wave pyramidal absorber-array structure studied in this work. For the above reasons, frequency-domain full-wave simulations are performed in this work using the RF Module of COMSOL Multiphysics [33]. The normal-incidence reflectivity is obtained by directly solving Maxwell's equations. To characterize the electromagnetic behavior of a large-scale array, the simulations employ a unit-cell model with Floquet periodic boundary conditions, while a perfectly matched layer (PML) is introduced along the propagation direction to absorb outgoing waves, thereby equivalently modeling the incidence and reflection processes of an infinite periodic array in free space. Meanwhile, to represent the suppression of electromagnetic transmission by the bottom OFHC backplate, the bottom surface of the structure is approximately modeled as a perfect electric conductor (PEC) boundary in the simulation. The material electromagnetic parameters are taken from the measurement results presented in Section 2.1.1. To ensure numerical accuracy, the mesh is determined according to the wavelength-based meshing criterion recommended by COMSOL [34]: within the simulated frequency range, the maximum surface mesh-element size of the absorber unit, $h_{\max}$, should not exceed one fifth of the local wavelength in the material, $\lambda_{\text{local}}$, where $\lambda_{\text{local}} = \lambda_0/\sqrt{\varepsilon'\mu'}$ ,namely, $h_{\max} \leq \lambda_0/5\sqrt{\varepsilon'\mu'}$ [29]. A schematic of the model is shown in Fig. 4B.

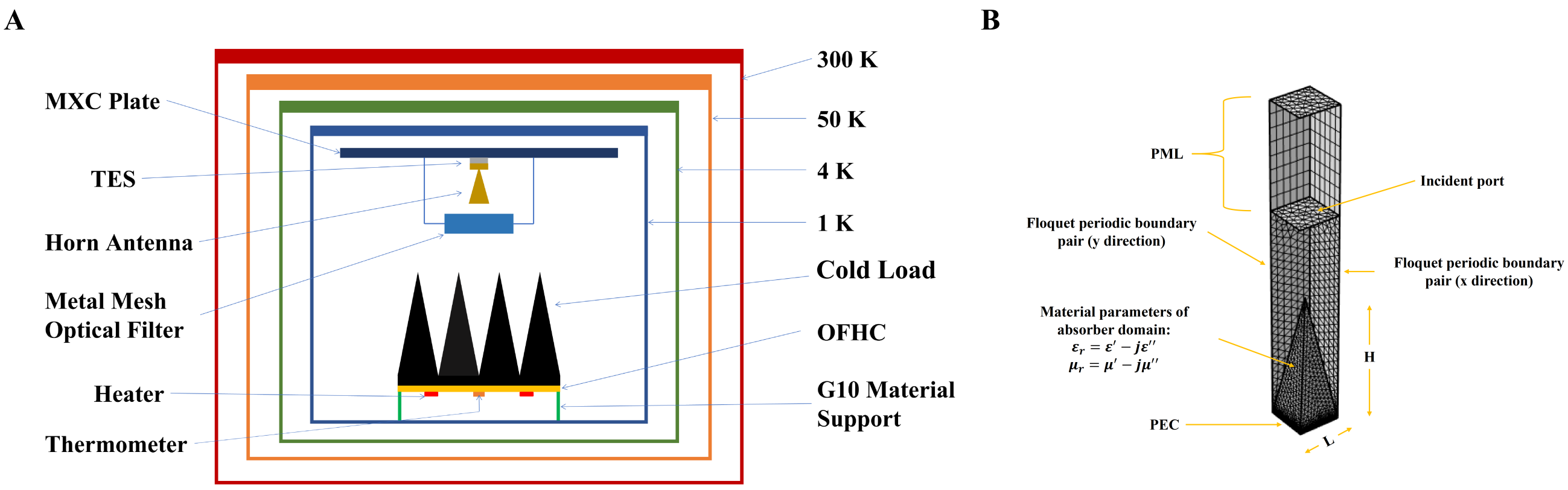

**Fig. 4: System integration and electromagnetic simulation model of the cold load absorber. (A) Schematic of the proposed assembly configuration of the cold load absorber in the DR. (B) Simulation model for electromagnetic simulation of the cold load absorber. The lateral faces are assigned Floquet periodic boundaries, and two opposite faces are hidden for clarity.**

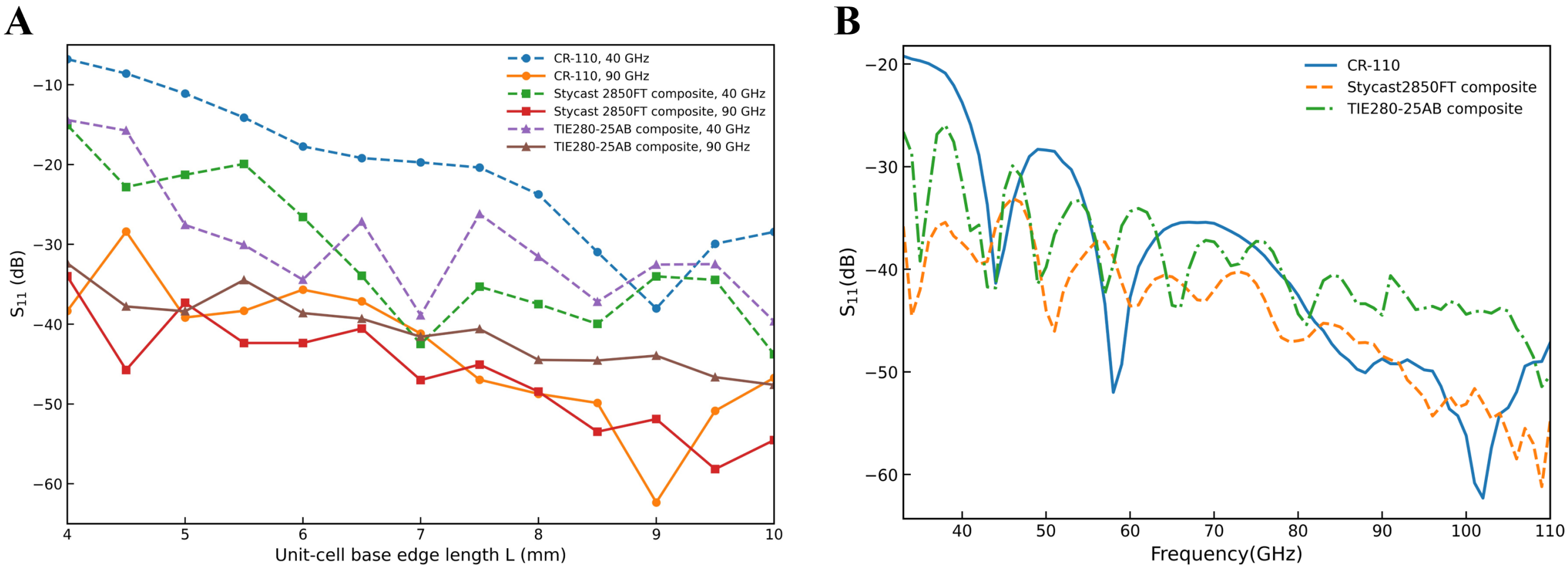

**Fig. 5: Structural optimization and broadband electromagnetic response of the cold load absorbers. (A) Simulated $S_{11}$ parameters as a function of the pyramid base side length $L$ at 40/90 GHz ($H = 3L$). (B) Simulated normal-incidence $S_{11}$ parameters of the optimized absorber structures over 33–110 GHz ($L = 8$ mm).**

Fig. 5A shows the simulated reflectance parameters of the absorber structures as a function of the absorber-pyramid unit size $L$ over the range of 4–10 mm. When $L$ exceeds 8 mm, the reflectance parameters of the absorber structures fabricated from all absorber materials remain below −20 dB at 40 and 90 GHz. Although larger dimensions can further reduce the reflectance parameters, they also occupy more space inside the DR. Therefore, the absorber-pyramid unit size is selected as $L = 8$ mm and $H = 24$ mm, and the overall lateral diameter of the absorber structure is limited to no more than 172 mm.

Fig. 5B further presents the simulated $S_{11}$ parameters of absorber structures based on CR-110, Stycast 2850FT composite, and TIE280-25AB composite under normal incidence over the 33–110 GHz frequency range. Simulation results show that all three material structures achieve low reflectance across the entire band, with $S_{11}$ values consistently below −20 dB and decreasing further with increasing frequency. This indicates that the pyramidal structure provides effective impedance grading and low normal-incidence reflection over a broad frequency range. All three structures satisfy the preset low-reflectance screening criterion near 40/90 GHz. To evaluate the influence of possible temperature-dependent changes in the dielectric properties, additional simulations were performed using different permittivity values. The results show only small changes in the simulated $S_{11}$ at 40/90 GHz. This indicates that the electromagnetic performance of the pyramidal array is not highly sensitive to moderate variations in the dielectric properties.

After the basic dimensions were determined, the $S_{11}$ parameters of the absorber structures with the three absorber materials were further calculated at 40/90 GHz as a function of the angle of incidence, as shown in Fig. 6. The simulation results show that, for the same structural dimensions, all material structures achieve $S_{11} < -20$ dB within the target frequency bands. At 90 GHz, the reflectance parameters are further reduced, approaching

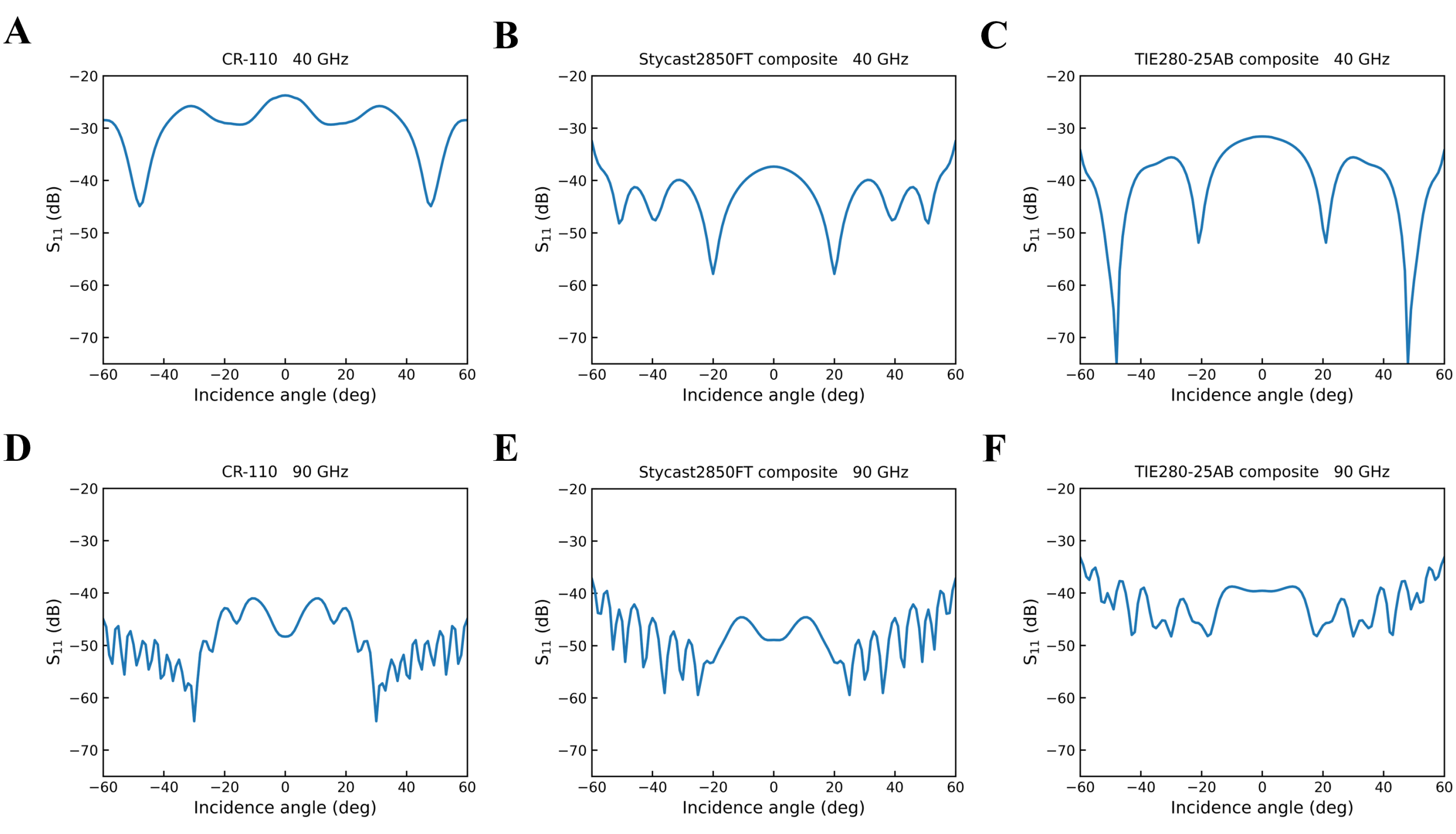

**Fig. 6: Dependence of the simulated $S_{11}$ parameters on incidence angle for the cold-load absorbers.**

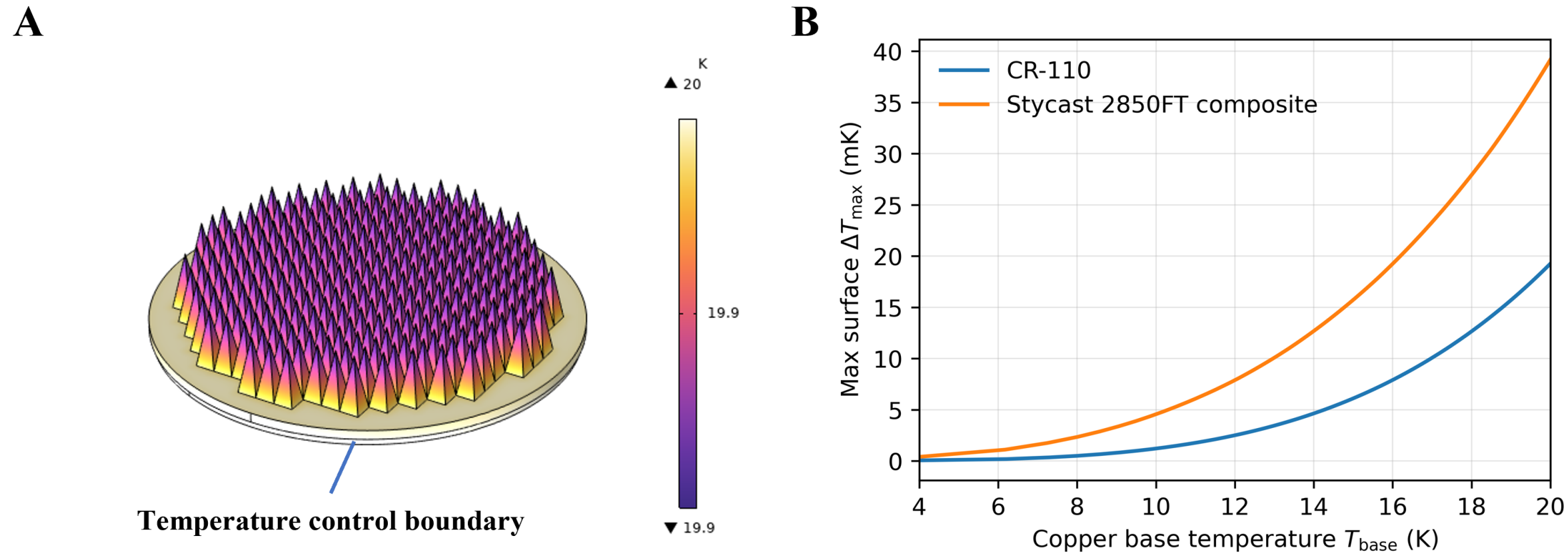

**Fig. 7: Simulated maximum surface temperature difference of the absorber structures over 4–20 K. (A) Simulated steady state temperature distribution of the absorber structures. (B) Maximum surface temperature difference $\Delta T_{\max}$ of the absorber structures as a function of operating temperature over 4–20 K.**

$S_{11} < -30$ dB, indicating low simulated reflection over the investigated incidence-angle range.

These results support selecting CR-110 and the Stycast 2850FT composite for subsequent low-temperature evaluation, while TIE280-25AB remains a preliminary material for further study.

### *2.2.2 Heat Transfer Simulation*

After determining the basic dimensions, we performed static steady-state temperature-field simulations using the Heat Transfer in Solids Module of COMSOL Multiphysics. The goal was to evaluate the temperature gradient of the cold load prototype at thermal equilibrium over 4–20 K. The boundary conditions were set as follows. The absorbing surface was assigned a surface-to-ambient radiation boundary with an emissivity of $\epsilon = 0.7$. The ambient temperature was set to 0.7 K, representing the typical operating temperature of the 1 K stage in the DR. All other external surfaces were treated as thermally insulated, approximating the adiabatic condition of a vacuum environment. The emissivity value of 0.7 was adopted as a conservative engineering estimate, accounting for the non-ideal broadband thermal-emission behavior of the absorbing surface at cryogenic temperatures. This thermal emissivity is used only in the heat-transfer calculation and is distinct from the millimeter-wave reflectivity evaluated by $S_{11}$.

For the material properties, the copper base was assigned the low-temperature thermophysical properties of RRR = 100 OFHC taken from the NIST database [35], while the low-temperature thermophysical parameters of the two composite absorbing materials are given in Section 2.1.2. In this simulation, a heating boundary was applied to the bottom surface of the copper base, and a parametric sweep was performed over the 4–20 K temperature range to extract the maximum temperature difference, $\Delta T_{\max}$, on the absorbing surface of the absorber structures at each temperature point.

As shown in Fig. 7, the $\Delta T_{\max}$ values of both material structures are below 40 mK within the target temperature range. This small predicted temperature difference supports the use of thermal simulation as a preliminary screening tool. The actual temperature distribution and transient response will be verified in future low-temperature experiments.

## 2.3 Fabrication of the absorber prototypes

We fabricated the absorber prototypes using a mature silicone replica-molding technique and further optimized the process to improve pyramid-tip filling, reduce bubble entrapment, and facilitate damage-free demolding [29]. First, an aluminum master mold with the designed pyramidal-array geometry was fabricated and placed in a polytetrafluoroethylene (PTFE) container to define the outer boundary of the mold and facilitate subsequent demolding. It was then used to prepare a silicone negative mold. The silicone mold provides sufficient flexibility for demolding while preserving the sharp pyramid tips. The absorber mixture was then prepared and degassed under vacuum to reduce trapped bubbles. The silicone mold was gradually filled with the absorber slurry rather than filled all at once, which helped improve the filling of the pyramid tips and reduce bubble entrapment in the grooves. After filling, an OFHC backplate was attached to the bottom of the absorber and served as both the mechanical support and thermal-control interface for heater and thermometer mounting. The assembled structure was cured under the selected curing conditions and then carefully demolded. The fabricated samples were inspected to confirm complete pyramid formation, good bonding to the copper backplate, and the absence of visible cracks or delamination. The fabrication flow is illustrated in Fig. 8.

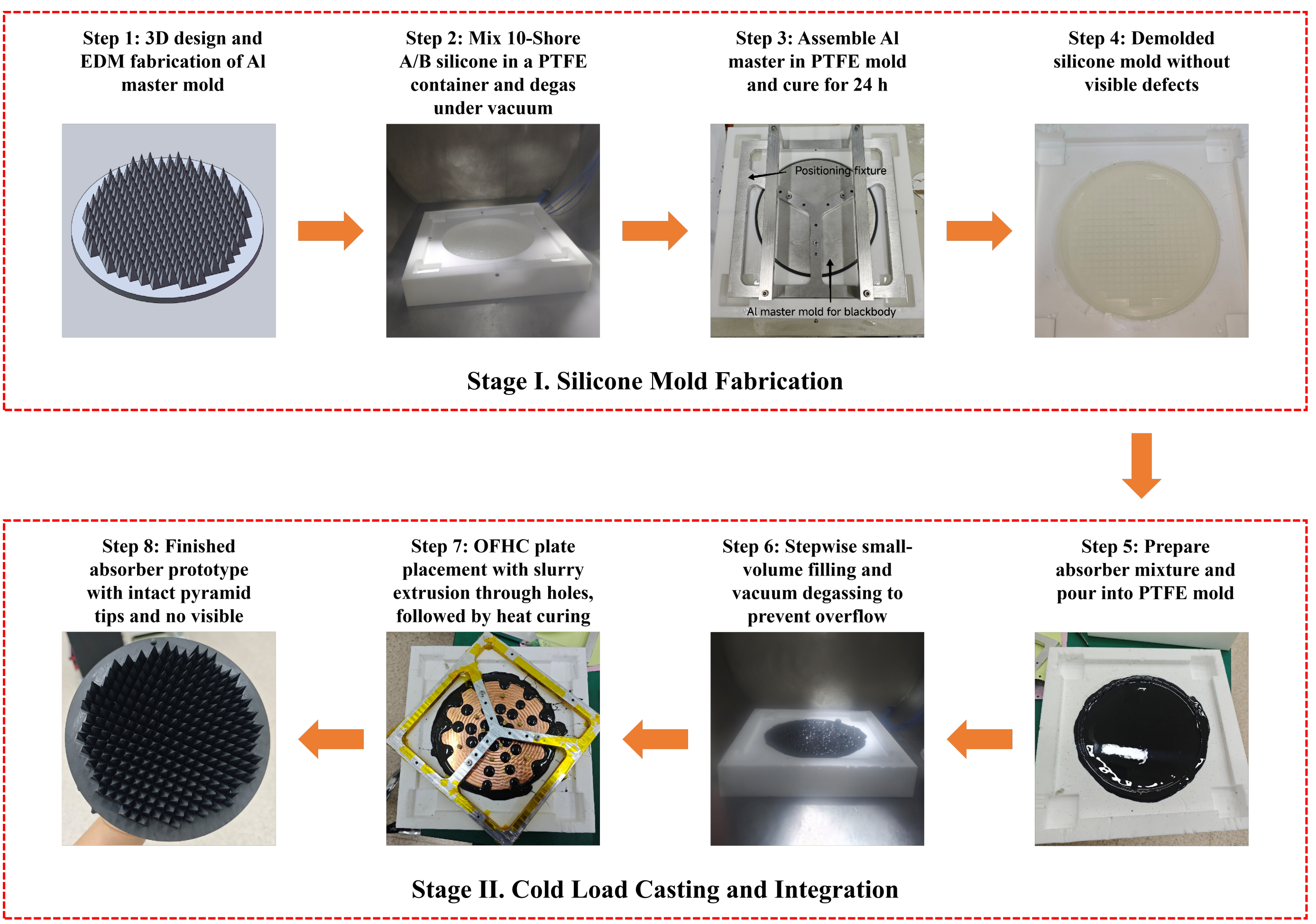

**Fig. 8: Fabrication process of the absorber prototype.**

CR-110 is a mature, commercially available castable microwave absorber, and its cryogenic reliability has been validated in multiple previous studies. The Stycast 2850FT composite, by contrast, is normally applied as a coating on metal surfaces [18, 36]. We therefore conducted repeated thermal-cycling tests between room temperature and 77 K to verify that its replicated molded form could withstand cryogenic conditions [24]. No cracking, delamination, or peeling was observed after cycling, supporting the fabrication feasibility of the prototype for subsequent low-temperature evaluation.

## 3. ROOM-TEMPERATURE REFLECTIVITY MEASUREMENT OF COLD-LOAD ABSORBERS

In the millimeter/submillimeter-wave band, CR-110- and Stycast 2850FT-based absorbers remain lossy at cryogenic temperatures, although their absorption loss may decrease with decreasing temperature [37]. Room-temperature reflectivity measurements have also been used for the preliminary evaluation of structured absorbers intended for cryogenic applications [38]. In this work, normal-incidence $S_{11}$ is used only as a preliminary screening metric for absorber structures. Because the OFHC backplate suppresses transmission, low total reflectance is a necessary condition for high absorptance. However, the measured $S_{11}$ cannot independently determine the absolute absorptance, emissivity, or radiation performance of the cold load. Therefore, a free-space VNA method was adopted to identify low-reflectance candidates for subsequent low-temperature testing.

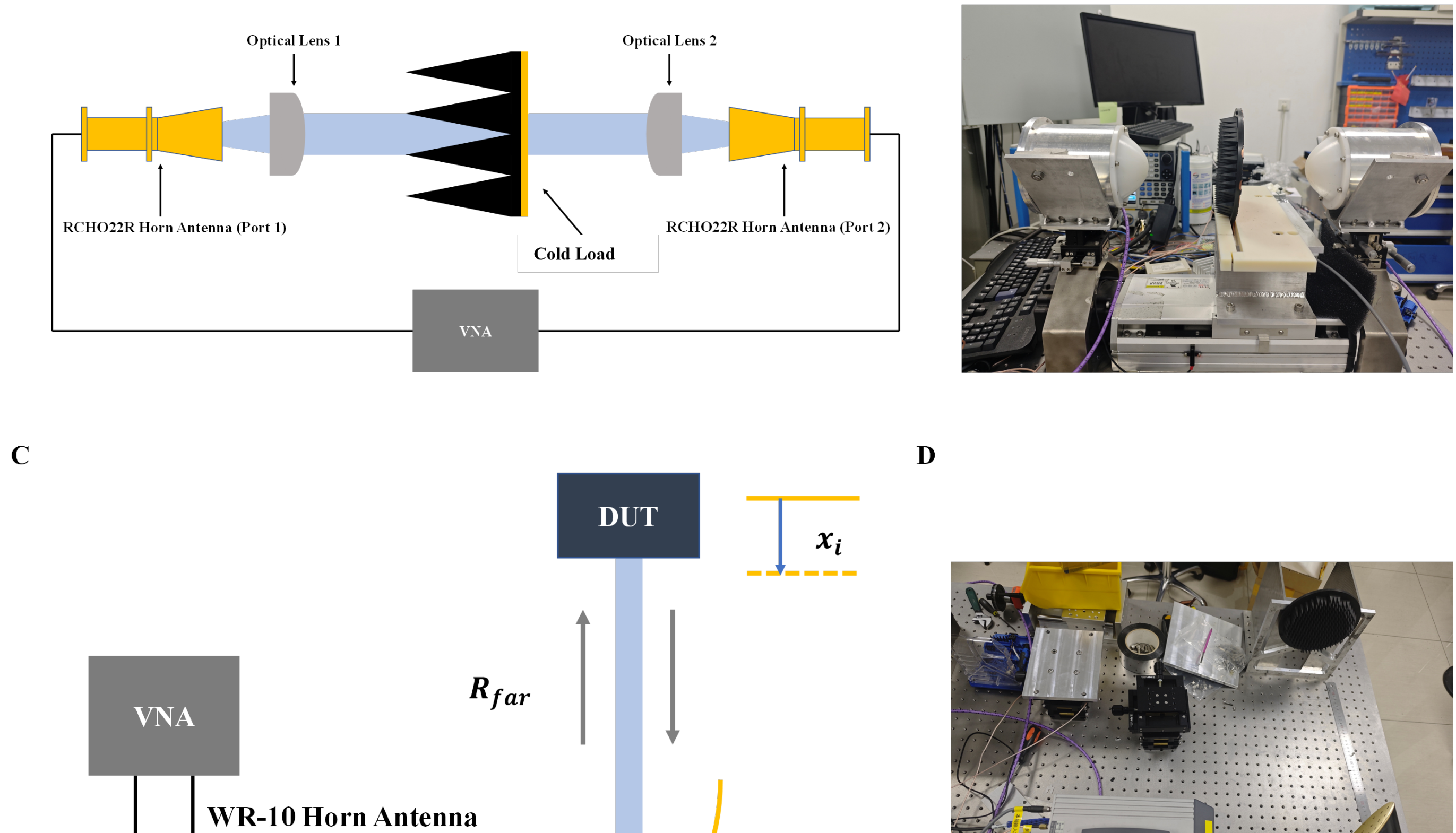

**Fig. 9: Measurement systems for the cold-load absorber reflectivity in the 33–50 GHz and 75–110 GHz bands. (A) Schematic of the 33–50 GHz measurement setup. (B) Photograph of the 33–50 GHz measurement setup. (C) Schematic of the 75–110 GHz measurement setup; (D) Photograph of the 75–110 GHz measurement setup.**

A dedicated cryogenic platform based on a 40 GHz TES detector is under construction. It will be used for low-temperature optical-power measurements of the cold load prototype.

## 3.1 Setup of the Measurement System

For the 33–50 GHz band, we used a two-port free-space measurement configuration [39]. System errors were removed by TRL (Thru–Reflect–Line) calibration [40]. Two RCHO22R horn antennas were arranged symmetrically with a separation of 225 mm, corresponding to about 30 wavelengths at the 40 GHz center frequency. This spacing satisfies the standard far-field criterion and reduces near-field effects and wavefront curvature. A collimating lens mounted in front of each horn converges the beam to a spot diameter of approximately 52.5 mm, or roughly 7 wavelengths at 40 GHz, matching the effective area of the absorber prototype. In total, 201 frequency points were sampled.

For the 75–110 GHz band, we developed an in-house free-space measurement system based on a multi-position calibration method derived from the TRL principle. During measurement, the device under test (DUT) is translated slightly along the surface normal to introduce a known optical path difference. The $S_{11}$ data acquired at multiple positions are then combined and solved by the least-squares method to separate the instrument response and recover the reflection coefficient of the DUT [41]. In this experiment, $x_i = 0$–1.6 mm with a step size of 0.2 mm, corresponding to a phase span of approximately 0.4–0.6$\lambda$, where $\lambda$ is the wavelength.

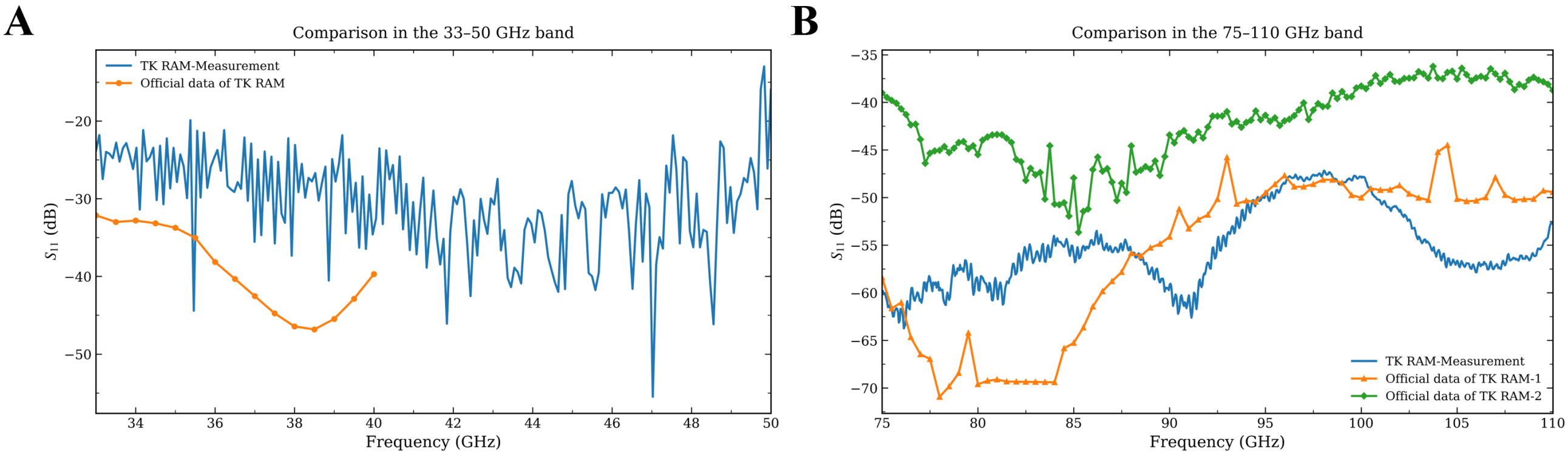

**Fig. 10: Comparison of the $S_{11}$ parameters of TK RAM under normal-incidence. (A) Comparison among the measured results and official data in the 33–50 GHz band. (B) Comparison among the measured results and two sets of official data in the 75–110 GHz band.**

The far-field distance between the antenna and the sample was determined from $R_{\text{far}} \geq 2D^2/\lambda$, where $D$ is the maximum linear dimension of the radiating aperture. For the pyramidal horn, $D$ was taken as the largest dimension of the horn aperture, giving a minimum far-field distance of $R_{\text{min}} \approx 370$ mm. An off-axis parabolic mirror (OAP) then collimates the horn radiation at its focal point. Using the nominal E-plane and H-plane half-power beamwidths (HPBWs) of the WR-10 horn, the collimated spot diameter was estimated to be much smaller than the 172 mm sample diameter, comfortably meeting the beam-coverage requirement. A total of 801 frequency points were sampled in this band.

The measurement systems are shown in Fig. 9.

To verify the reliability of the two free-space reflectivity measurement systems and the associated calibration methods, a commercial absorbing material, TK RAM, was selected as a reference sample for comparative study [42], as shown in Fig. 10.

It should be noted that the official data were not obtained under exactly the same measurement conditions as those used in this work. In addition, certain discrepancies may arise from sample-to-sample variations, while the digitization of curves from the official website may also introduce additional errors. Therefore, some deviations among the datasets in point-by-point values and in the positions of local minima are understandable. Nevertheless, the reflectivity levels of the different datasets within the corresponding frequency bands remain generally within the same or adjacent orders of magnitude, with no substantial discrepancy in magnitude, and they exhibit similar overall frequency-dependent trends. These results indicate that the free-space measurement systems established in this work, together with the adopted calibration methods, can accurately characterize the reflectivity of samples in the millimeter-wave band.

## 3.2 Measured Results

To evaluate the low-reflectance performance of the absorber structures, the measured $S_{11}$ values were compared with the full-wave simulation results. Fig. 11A– 11B compares the simulated and measured results in the frequency band centered at 40 GHz. For both absorber prototypes, the measured $S_{11}$ values remain below $-20$ dB, satisfying the preset room-temperature low-reflectance screening criterion. In the CR-110 prototype, noticeable discrepancies are observed between simulation and measurement in both the positions of local minima and the reflection depths. These deviations arise mainly from fabrication tolerances, assembly gaps, edge diffraction in the free-space measurement setup, and residual calibration errors [43]. The Stycast 2850FT composite prototype, however, exhibits a reflectance level comparable to that of the commercial TK RAM absorber, demonstrating that it satisfies the low-reflectance design requirement near 40 GHz and supports its use in the cold load prototype for subsequent cryogenic testing. Fig. 11C further presents the results in the frequency band centered at 90 GHz. The $S_{11}$ values of the CR-110 absorber structure are lower than $-30$ dB. For the Stycast 2850FT composite absorber structure, the $S_{11}$ values are lower than −40 dB at most frequency points, with a minimum value reaching −66 dB, indicating strong suppression of normal-incidence return reflection in the 90 GHz band. Overall, the low-reflectance performance of the Stycast 2850FT structure near 90 GHz is comparable to that of TK RAM, and it even exhibits lower reflection at some frequency points.

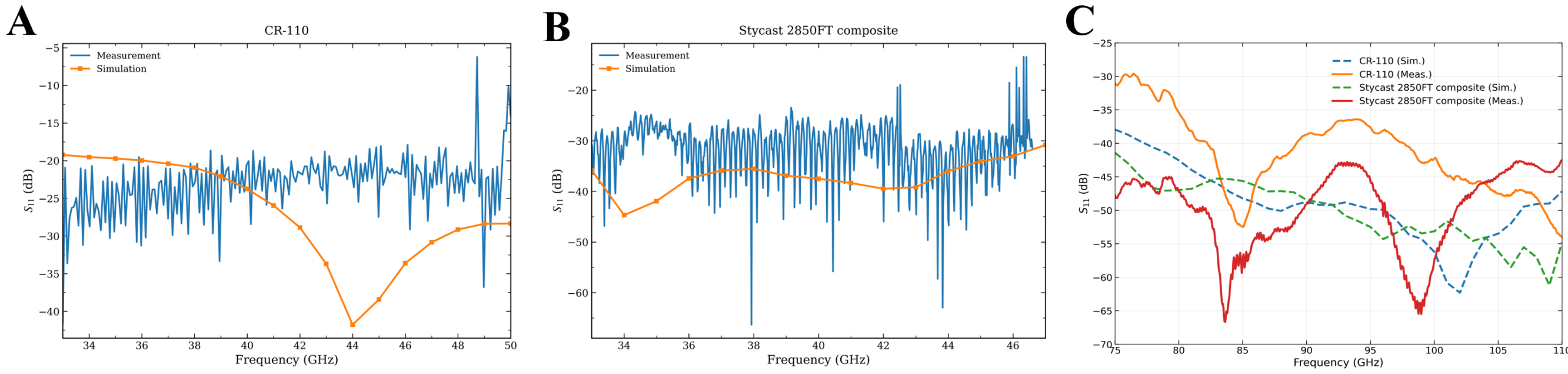

**Fig. 11: Simulated and measured normal-incidence $S_{11}$ parameters of the fabricated absorber prototypes in the target frequency bands. (A) Comparison between simulation and measurement for the CR-110 absorber prototype in the frequency band centered at 40 GHz. (B) Comparison between simulation and measurement for the Stycast 2850FT composite absorber prototype in the frequency band centered at 40 GHz. (C) Comparison between simulation and measurement for the two absorber prototypes in the frequency band centered at 90 GHz.**

The electromagnetic simulations in this work serve two primary purposes: to guide parameter optimization in the structural design of the absorber structure, and to verify the feasibility of achieving low reflectance in the target frequency bands. Exact point-by-point agreement between the simulated and measured curves is neither expected nor required. Although some discrepancies remain, both in the values at specific frequency points and in the positions of local minima, the two sets of results agree on the essential point: the designed absorber structure meets the $S_{11} < -20$ dB requirement across the target operating bands. These results show that the model can guide the low-reflectance design, while the cryogenic radiometric performance will be evaluated using the 40 GHz TES-based platform in future work.

## 4. SUMMARIES

This work presents the development and preliminary characterization of a variable-temperature millimeter-wave cold load prototype intended for future 40/90 GHz TES calibration. Focusing on absorber materials, simulation, fabrication, and testing, this study investigates CR-110 and the Stycast 2850FT composite through electromagnetic characterization, thermal-feasibility analysis, full-wave and steady-state simulations, prototype fabrication, and room-temperature reflectivity measurements. TIE280-25AB was also preliminarily evaluated as an alternative material.

The simulations indicate that the CR-110 and Stycast 2850FT composite structures exhibit low reflectance and small predicted steady-state temperature gradients over 4–20 K. The measured $S_{11}$ values satisfy the preset room-temperature low-reflectance screening criterion and generally agree with the simulations. The Stycast 2850FT composite exhibits particularly low reflection near 90 GHz, with a minimum $S_{11}$ of −66 dB, and its reflectivity is comparable to that of commercial TK RAM. These results support selecting both structures for subsequent low-temperature evaluation.

The measured electromagnetic parameters and RF simulations also indicate promising low-reflectance performance for the TIE280-25AB composite over 33–110 GHz. However, its low-temperature thermal properties, prototype fabrication, and broadband reflectivity measurements remain to be investigated before it can be directly compared with the other two structures.

In summary, the measured and simulated results show that the proposed prototype is a promising candidate for a cryogenic millimeter-wave blackbody source. Further low-temperature radiometric and TES-based optical-power measurements are required to verify its final performance.

## ACKNOWLEDGEMENTS

The authors gratefully acknowledge Hefei Zhileng Cryogenic Technology Co., Ltd. for its guidance and technical support in the heat transfer simulation of this work, and Tianjin SiDaRui Technology Development Co., Ltd.

for its assistance in the CNC machining of the absorbing-material samples. Their support contributed significantly to the simulation, measurement, and fabrication presented in this work. This work was supported by the National Key Research and Development Program of China (Grant No. 2021YFC2203400) and the Youth Innovation Promotion Association of CAS (Grant No. 2021011). The authors also thank all organizations and individuals who supported this research.

## AUTHOR CONTRIBUTIONS

Ruya Cong, Zhengwei Li, and Yu Xu conceived the overall research plan and developed the technical concept of the millimeter-wave cold load prototype. Junjie Zhou designed the multiphysics simulation scheme. Jianrong Cai carried out the structural design and explored the fabrication process. Haoxuan Gao and Kang Zhang prepared the samples. Xuefeng Lu established the $S_{11}$ measurement system, performed the reflectivity measurements of the absorber prototypes, and provided access to the licensed COMSOL Multiphysics software used in this work. Xingyuan Hou provided guidance on the electromagnetic characterization of the materials, and Shaoliang Wang provided guidance on the DR engineering. All authors read and approved the final manuscript.

## DECLARATION OF INTERESTS

The authors declare no competing interests.